\documentclass[]{piparticle-final}
\usepackage{graphicx}
\usepackage{amsmath}

\usepackage{cite} 
\usepackage{epstopdf}

\begin{document}

\volume{6}               
\articlenumber{060007}   
\journalyear{2014}       
\editor{L. A. Pugnaloni}   
\reviewers{R. Ar\'evalo,  School of Physical and Mathematical Sciences, \\\mbox{ }\hspace{3.5cm} Nanyang Technological University, Singapore.
}  
\received{14 June 2014}     
\accepted{7 October 2014}   
\runningauthor{C. F. M. Magalh\~aes \itshape{et al.}}  
\doi{060007}         

\title{Jamming transition in a two-dimensional open granular pile with rolling resistance}

\author{C. F. M. Magalh\~aes,\cite{unifei}\thanks{E-mail: cfmm@unifei.edu.br}\hspace{0.5em}
        A. P. F. Atman,\cite{cefetmg,inctsc}\thanks{E-mail: atman@dppg.cefetmg.br} \hspace{0.5em}
        G. Combe,\cite{ujf}\hspace{0.5em}
        J. G. Moreira\cite{ufmg}\thanks{E-mail: jmoreira@fisica.ufmg.br}}

\pipabstract{
We present a molecular dynamics study of the jamming/unjamming transition in two-dimensional granular piles with open boundaries. The grains are modeled by viscoelastic forces, Coulomb friction and resistance to rolling. Two models for the rolling resistance interaction were assessed:  one considers a constant rolling friction coefficient, and the other one a strain dependent coefficient. The piles are grown on a finite size substrate and subsequently discharged through an orifice opened at the center of the substrate. Varying the orifice width and taking the final height of the pile after the discharge as the order parameter, one can devise a transition from a jammed regime (when the grain flux is always clogged by an arch) to a catastrophic regime, in which the pile is completely destroyed by an avalanche as large as the system size. A finite size analysis shows that there is a finite orifice width associated with the threshold for the 
unjamming transition, no matter the model used for the microscopic 
interactions. As expected, the value of this threshold width increases when rolling resistance is considered, and it depends on the model used for the rolling friction.
}

\maketitle

\blfootnote{
\begin{theaffiliation}{99}

   \institution{unifei} Universidade Federal de Itajub\'a - Campus de Itabira, Rua Irm\~a Ivone Drumond, 200, 35900-000 Itabira, Brazil.
   
   \institution{cefetmg} Departamento de F\'{i}sica e Matem\'atica, Centro Federal de Educa\c{c}\~ao Tecnol\'ogica de Minas Gerais, Av. Amazonas, 7675, 30510-000 Belo Horizonte, Brazil.
   
   \institution{inctsc} Instituto Nacional de Ci\^encia e Tecnologia Sistemas Complexos, 30510-000 Belo Horizonte, Brasil.
   
   \institution{ufmg} Universidade Federal de Minas Gerais, Caixa Postal 702, 30161-970 Belo Horizonte, Brasil.
   
   \institution{ujf} UJF-Grenoble 1, Grenoble-INP, CNRS UMR 5521, 3SR Lab. Grenoble F-38041, France.   

\end{theaffiliation}
}

\section{Introduction}

Granular materials are ubiquitous either in nature ---desert dunes, beach sand, soil, etc.--- or in industrial processes as mineral extraction and 
processing, or food, construction and pharmaceutical industries \cite{antony,duran,halsey}. In fact, any particulate matter made of macroscopic 
solid elements can be classified as granular material. The vast phenomenology exhibited by these systems combined with an incomplete understanding 
about the microscopic physical mechanisms responsible for the macroscopic behavior of these materials have motivated the increasing interest of the 
physics community in the past years \cite{yu,midi}.

Although materials of this class are not sensitive to thermal fluctuations, they can be found at gas, liquid or solid phases \cite{jaeger}. 
The transition between solid and liquid phases in granular matter, which is commonly referred to as jamming/unjamming transition, has been 
extensively studied from both theoretical and experimental perspectives \cite{yu,jaeger,cixous,atman13,liu,majmudar,mankoc}. Currently, a great effort is being made to 
understand the nature of this transition, which is still a subject of debate \cite{atman14}. The jamming/unjamming transition is not a 
specific property of granular matter, being observed in many kinds of materials, such as foams \cite{katgert}, emulsions \cite{denkov}, colloids 
\cite{kumar}, gels \cite{fluerasu}, and also in usual molecular liquids \cite{duplantier} ---glass transition. Liu and Nagel \cite{liu} proposed a 
general phase diagram as an attempt to unify the several approaches to study jamming/unjamming in disordered materials. This work has motivated 
several theoretical, experimental and numerical investigations, but a comprehensive understanding of this transition is still lacking.

O'Hern \textit{et al.} \cite{hern1,hern2} have performed numerical simulations of granular materials approaching to jamming in two and three 
dimensions. They have explored the packing fraction axis of the general phase diagram proposed by Liu and Nagel and have demonstrated, by means 
of finite-size analysis, the existence of a unique critical point in which the system jams in the thermodynamic limit. The authors have also shown 
some evidence that this point is an ordinary critical point, indicating that the jamming transition would be a second order phase transition. These 
results were corroborated later by experiments (c.f. Majmudar \textit{et al.} \cite{majmudar}) and by simulations (c.f. Manna and Khakhar \cite{manna}). 
In Ref. \cite{manna}, the authors have revealed evidence of self-organized criticality (SOC) by measuring the internal avalanches resulting from the opening of 
an orifice at the bottom of granular piles.

Experimental investigation of the jamming transition in granular materials under gravitational field has been conducted in a variety of ways, addressing 
the role of many parameters, like the grain shape, the friction coefficient, and the system geometry. However, a common feature of all these approaches is the 
analysis of the granular flow through bottlenecks. 

The jamming of three-dimensional piles seems to be settled after the work of Zuriguel \textit{et al.} \cite{zuriguel}. They have demonstrated experimentally,
for piles composed of different kinds of grains, the divergence on the mean internal avalanche size. It means that, as the outlet size approaches a critical 
value, the internal avalanche increases without limit and a permanent flow is established. This critical outlet is insensitive to the density, stiffness and 
roughness of the grains, but shows a significant dependence on the grain shape. For spherical grains, a critical outlet width $w_c \sim 4.94 d$ was 
obtained, for cylindrical grains $w_c \sim 5.03 d$ and for rice grains $w_c \sim 6.15 d$.

Nevertheless, the jamming transition in two-dimensional	 piles is still a question under debate. In order to address it, To \textit{et al.} \cite{to} 
have carried out experiments using two-dimensional hoppers in order to find a critical outlet size for jamming events. The jamming probability $J$ has 
presented a rapid decay from $J = 1$ to $J = 0$ close to the aperture width $w \sim 3.8 d$, signaling a possible phase transition. The authors discussed, 
based on a restricted random walk model, the connection between jamming and the arch formation mechanism. Nevertheless, the point needs further investigation, especially at the limit of high hopper angle. There exist several works \cite{garci,drescher,maga2} focused on the mechanisms of arch formation, but 
none had explored its relation to jamming probability.

Janda \textit{et al.} \cite{janda} have made some progress by simulating discharges of two-dimensional silos. They have improved the definition of 
jamming so that the internal avalanche size is considered. This modification addresses the extremely long relaxation times associated with jamming. 
Within the framework of a probabilistic theory concerning the arch formation, the authors have tested two hypothesis for the internal avalanche behavior: 
a functional form that predicts a divergence in the mean size, and a functional form where the mean size exists for all values of the orifice width $w$. 
Since the latter one is more compatible with the arch formation model, the authors claimed that ``no critical opening size exists beyond which there is 
not jamming" \cite{janda}.

The results mentioned so far are related to fully confined systems. Recently, a simulation study on discharges of granular piles with open 
boundaries \cite{maga1} provided new insights on the problem. The piles are composed by homogeneous disks interacting via elastic and frictional 
forces. Using finite size analysis, the authors have shown that a catastrophic regime, in which the pile is completely destroyed by the opening of 
the orifice, is well defined. At the limit of infinitely large systems, the catastrophic regime coincides with the unjammed phase, since it implies a 
divergent internal avalanche. Hence, the results indicate the existence of the jamming transition. It is important to note, however, that the pile 
geometry could probably play a role in the causes for this distinct behavior, due to the absence of the Janssen effect, but further investigations 
are necessary to confirm this point.

In the present work, the investigation of jamming in 2D open systems is extended in order to consider a rolling resistance term in the grain interaction 
model, following the prescription adopted by Chevalier \textit{et al.} \cite{chevalier}. The main objective of this study is the verification of rolling 
resistance influence on jamming. Many factors contribute to the appearance of rolling friction, including microsliding, plastic deformation, surface adhesion, 
grain shape, etc., but mainly, it is due to the contact deformation \cite{ai}. Here, it will be taken into account only the effect due to the contact 
deformation by implementing the micromechanical model proposed by Jiang \textit{et al.} \cite{jiang}. The rolling friction produces a resistance to roll 
which, among other effects, is responsible for granting more stability to granular piles \cite{zhou}, and for the occurrence of different types of failure 
modes in granular matter \cite{iwashita,li}. Rolling friction was also used to model a system of polygonal grains by making a correspondence between the 
rolling stiffness and the number of sides of the polygon \cite{estrada}. It was demonstrated that it is an essential ingredient to reproduce experimental 
compression tests in mixtures of two-dimensional circular and rectangular grains \cite{charalampidou}. These facts suggest that jamming could be affected 
by rolling friction. Nevertheless, most studies on granular materials based on computer simulations do not deal with it.

The paper is organized as follows: after a review in the Introduction, the next section is concerned with the methodology. Then, we present 
the results and a brief discussion. Finally, the last section gathers the conclusions and some perspectives.

\section{Methods}

The jamming transition is assessed by means of discharges of granular piles, simulated using the molecular dynamics method \cite{rapaport} with the 
Velocity-Verlet algorithm \cite{swope}. In a few words, the molecular dynamics consists in integrating numerically the equations of motion that governs the 
system dynamics. The system is constituted by N free grains governed by Newton's second law and by a finite and horizontally aligned substrate made of fixed 
grains. The free grains, which will form the pile, are homogeneous bi-dimensional disks that are free to translate and rotate around their center, and whose 
radii are uniformly distributed around an average value $d$, a small polydispersity of 5\% was imposed in order to avoid crystallization effects. All spatial 
quantities will be expressed in terms of $d$. Since the grains have all the same mass density, their masses are proportional to their respective areas. 
Normalized by the mass of the heaviest grain, the masses are given by $m_i = d_i^2 / d_{max}^2$, where $d_{max}$ stands for the diameter of the largest 
grain. The finite substrate of length L is composed by fixed grains of the same kind but with a smaller and fixed diameter $d_s = 0.1 d$. These grains 
are aligned horizontally and are equally spaced in order to form a grid without gaps.

\begin{figure}
\begin{center}
\includegraphics[width=0.45\textwidth]{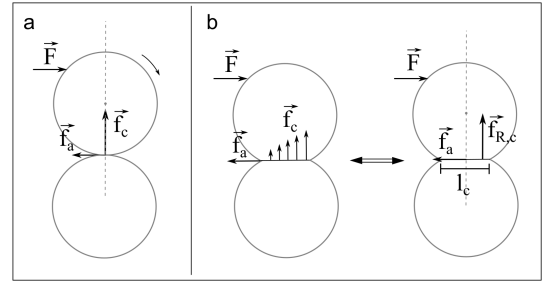}
\end{center}
\caption{In panel (a), a perfectly rigid grain rolling over another one is exhibited. The contact force is concentrated on one point and is aligned 
through the grain center, thus, not generating momentum with respect to the center. Panel (b) shows the same situation but with deformable grains. Now, 
the contact force is non-uniformly distributed over a segment of length $l_c$ (contact length). The resultant force $\vec F_{R,c}$ is dislocated from the 
center line and an opposing moment with respect to the center appears. 
\label{mrr}}
\end{figure}

The grains are subject to a uniform gravitational field orthogonal to the substrate line, and to short range binary interactions. Besides the visco-elastic 
and coulomb friction interactions used in past works \cite{maga2,maga1,goldenberg,pinto,atman09,cundall,mindlin}, two grains in contact are also subject to 
a rolling resistance moment due to the finite contact length $l_c$. The rolling resistance is introduced through a micromechanical model of the contact line 
between two grains \cite{jiang}. This model treats the contact as an object formed by a set of springs and dashpots connecting the borders of the two grains. 
As one grain rolls over the other, the springs in one side of the contact line contract while the springs in the opposite side stretch. This configuration 
generates an unbalanced force distribution and a consequent moment with respect to the grain center (see Fig. \ref{mrr}). This moment grows linearly as 
the grain rolls, until the rolling displacement $\delta_r$ reaches some threshold value at which the springs located near one end of the contact line break 
up and new ones emerge at the other end. At that time, the moment saturates at some value that depends on the properties of the grain. The rolling 
displacement is defined by $\delta_r = \sum ( \omega_i - \omega_j ) \Delta t$, where $\omega_i$ refers to the angular velocity of grain $i$, $\Delta t$ is 
molecular dynamics time step, and the sum runs over time during the whole existence of contact. Based on these assumptions, the authors have derived an 
analytic expression for the rolling resistance moment as a function of rolling displacement. They have also proposed a simplified version - the one used 
in this study - in order to improve numerical computations:

\begin{align}
\tau_{rr} = \begin{cases} - k_r \delta_r , & k_r |\delta_r| \leq \mu_r f_{el} \\
- \dfrac{\delta_r}{|\delta_r|} \mu_r f_{el} , & k_r |\delta_r| > \mu_r f_{el} ~ ,
\end{cases}
\end{align}
where $k_r$ is the rolling stiffness, $\mu_r$ is the coefficient of rolling resistance, and $f_{el}$ is the compressive elastic force, normal to contact 
line. As can be noted, the expression for the rolling resistance momentum possesses a striking resemblance with the Coulomb static friction force, and as 
a matter of fact, the rolling resistance interaction is implemented in the molecular dynamics algorithm in the same way as the static friction. The model 
predicts that the rolling stiffness and the coefficient of rolling resistance are related to the contact length $l_c$ by the equations $k_r = k_n l_c^2$ 
and $\mu_r = \mu l_c$, where $k_n$ and $\mu$ are respectively the normal elastic constant and friction coefficient. The values used for these parameters 
were the same as in Ref. \cite{maga1}, $\mu =0.5$ and $k_n =1000$ in normalized unities (see \cite{atman09} for further details). Two cases were 
investigated: systems in which the rolling resistance parameters $k_r$ and $\mu_r$ vary according to the above-mentioned equations, and systems with 
fixed rolling resistance parameters, assuming that all contacts have the same deformation value.

The simulation procedure consists of two steps: (1) the formation of the granular pile with open boundaries, by deposing grains from rest, under gravity, 
over the substrate until an stationary state is reached; (2) the discharge itself, which consists in opening an orifice of a given width at the center of the 
substrate. In the first step, the initial positions of the free grains are randomly sorted along a horizontal line located at height $L$ from the substrate - 
the releasing height is equal to the substrate length. To avoid initial overlapping of grains, a $50\%$ filling ratio was imposed to each line of grains 
released, and the time interval between successive rows is the inverse of the frequency $f$. Each row was released after the predecessor had fallen a 
distance equivalent to the maximum grain diameter. This deposition protocol mimics a dense rain of grains. During deposition, the grains may leave the 
system through the lateral boundaries, so that the total number of grains in the pile fluctuates as the process evolves. The release of grains ceases 
when the number of grains in the pile reaches a stationary value. The deposition phase ends only when a mechanical equilibrium state is attained, and the 
configuration is recorded for later analysis. The equilibrium state must satisfy the following criteria: mechanical stability, absence of slipping contacts, 
vertical and horizontal force balance, and vanishingly small kinetic energy \cite{atman05}. In the second step, the configuration recorded is loaded and an 
outlet of width $w$ is opened at the center of the substrate, allowing the grains to flow through it. As the grains pass through the outlet, they are removed 
from the system, in order to improve computational efficiency. The simulation runs until a new equilibrium state is reached, which happens either due to 
the formation of an arch above the outlet or after the pile has been completely discharged. The remaining pile configuration is then recorded.

The average height $h$ of the resulting pile after discharge, measured from the center of the substrate, was taken as an order parameter to distinguish 
between the two regimes. If $h$ does not change significantly with respect to the original height, it means that an arch does readily clog the flux after 
the orifice opening, but if $h = 0$, it means that the pile has collapsed, and a jammed state was not attained for the corresponding orifice diameter and 
substrate length.  As the orifice width $w$ approaches a certain threshold value $w_t$, the system suffers a transition from jammed to unjammed state. 
This threshold is defined as the orifice width for which the height fluctuations is maximum. The connection to the jamming transition occurs 
when the substrate length diverges since the collapse of an infinite substrate pile implies a continuous flowing state. Then, the jamming transition 
can be characterized by a critical aperture width, as defined by the following expression

\begin{align}
w_c = \lim_{L \rightarrow \infty} w_t(L) ~ .
\end{align}

\begin{figure}
\begin{center}
\includegraphics[width=0.2\textwidth]{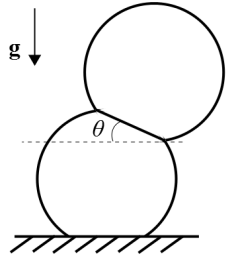}
\end{center}
\caption{Diagram of a possible equilibrium configuration if rolling resistance is considered. The grain on top has only one contact and, even so, it can sustain 
a stable position. The maximum value of the angle $\theta$ depends on the friction coefficient and on the rolling resistance parameters.} \label{z1eq}
\end{figure}

\begin{figure}
\begin{center}
\includegraphics[width=0.45\textwidth]{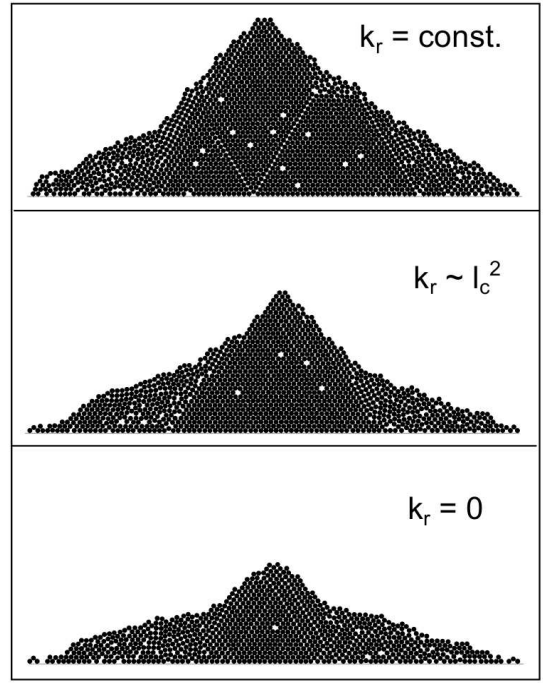}
\end{center}
\caption{Images of the pile equilibrium states for the three grain interaction models before the discharge step. The piles were grown over a $L = 100 d$ 
substrate and are representative samples from each type of pile. As shown in the figure, the top image represents a pile composed by grains with a fixed 
$k_r$ rolling resistance interaction, the image in the middle is a pile of grains with a $k_r \sim l_c^2$ rolling resistance interaction, and the bottom 
image is a pile of grains without rolling resistance.} \label{piles}
\end{figure}

\section{Results and Discussion}

Two different models of rolling resistance interaction were tested in the simulations: the original model proposed by Jiang \textit{et al.} \cite{jiang} 
mentioned earlier, with a rolling constant and a coefficient of rolling resistance that depends on the contact length 
($k_r = k_n l_c^2$ and $\mu_r = \mu l_c$)
and a simpler derived version, in which these parameters are constant over time assuming that $l_c = 0.05~d$ for all contacts. This fixed
$l_c$ model assumes a mean contact length equivalent to $5\%$ of the average grain diameter, a scenario which could be associated to a system composed by 
polygonal grains, for example, a extreme rolling resistance regime.

\begin{figure}[th]
\begin{center}
\includegraphics[width=0.68\columnwidth,angle=-90]{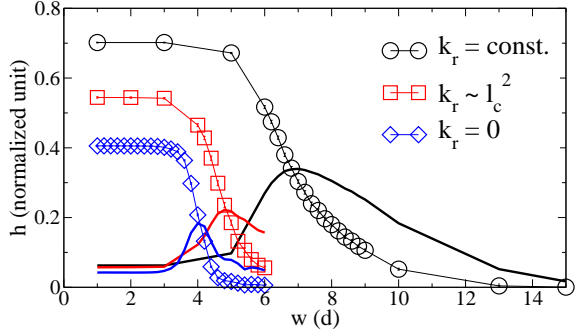}
\end{center}
\caption{Order parameter $h$ as a function of the orifice width $w$ for all types of pile. The graphs were generated from the simulation data of 
$L = 100 d$ piles. While the symbols represent the parameter $h$ itself, the lines indicate the corresponding fluctuations. The thick line is related 
to the fixed $k_r$ curve, the medium thickness line to the $k_r \sim lc2$ curve, and the dashed line to $k_r = 0$ curve.} \label{hxw}
\end{figure}

\begin{figure}[th]
\begin{center}
\includegraphics[width=0.65\columnwidth,angle=-90]{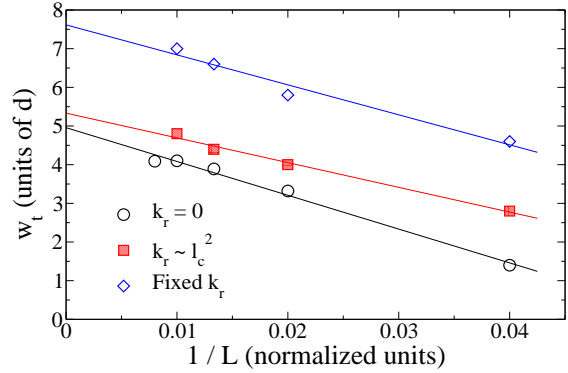}
\end{center}
\caption{Graphs of the threshold orifice width $w_t$ as a function of the reciprocal system size $1/L$ for models with and without rolling resistance. 
As the caption indicates, the symbols represent the data obtained from numerical simulations and the lines are fitting curves, which provide the respective 
values of $w_c$.} \label{wc}
\end{figure}

For either models, the numerical simulations described in the last section were carried out for various system sizes and orifice widths. Figure \ref{piles} 
shows equilibrium configurations of typical piles with $L = 100 d$ for the two tested rolling resistance models and for the absence of rolling resistance case. 
It can be seen that the inclusion of the rolling resistance term modifies significantly the macroscopic features of the pile. It provides more stability 
to the structure, which is reflected by a steeper free surface, a fact also observed elsewhere \cite{zhou}. Indeed, it should be noted that the rolling 
resistance makes possible some otherwise very unstable two-grain configuration, as exemplified in Fig.~\ref{z1eq}.

\begin{figure}
\begin{center}
\includegraphics[width=0.85\columnwidth,angle=-90]{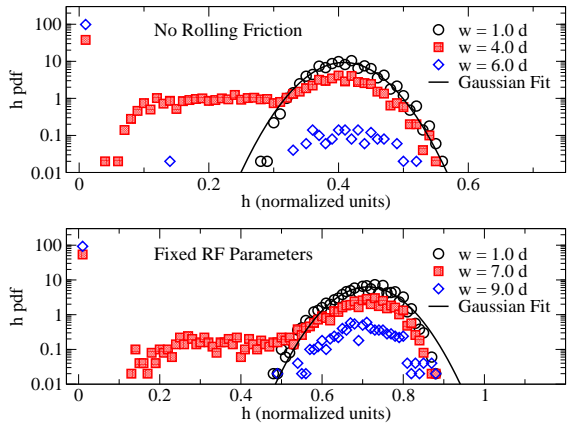}
\end{center}
\caption{Probability density functions of the order parameter $h$ for different values of $w$ in the case of absent rolling friction grains (top) and fixed 
rolling friction grains (bottom).} \label{hpdf}
\end{figure}

The behavior of the order parameter $h$ as function of $w$ is presented in Fig. \ref{hxw} for the three cases. Note that the transition region translates 
to the right as the rolling resistance becomes more important, which is an expected result since the increasing of stability allows the formation of larger 
arches. Figure \ref{wc} exhibits the dependence of $w_t$ (fluctuation maximum) on the system size for the three models considered and, again, the curves are 
dislocated along the $w_t$ axis. Nevertheless, they all share the same functional aspect, which is an evidence that the rolling 
resistance does not change the nature of the jamming transition, only the critical value of the threshold width. 
Despite the tendency to a heavy tail distribution observed for large $L$ and $w$, in all scenarios, the data 
suggest the existence of the jamming transition. It means that the features described in Ref. \cite{maga1} to characterize the transition were also
observed in both scenarios tested for the rolling resistance. The fitting values were $w_c = (5.3 \pm 0.1) d$ for contact length dependent $k_r$ model and 
$w_c = (7.6 \pm 0.2) d$ for the fixed $k_r$ one, while for the model without rolling resistance, $w_c = (5.0 \pm 0.1) d$ \cite{maga1}.
These results indicate that the rolling resistance only slightly changes the critical width when included in a system of disks, expressed
by the contact length model. But, in other approaches, as in the fixed length model, it can alter significantly the critical aperture width.

The probability density functions (PDF) of $h$ for the absent rolling friction and for the fixed rolling friction models are exhibited in Fig. \ref{hpdf}. 
For each model, the PDFs are obtained for three characteristic values of $w$: $w = 1.0 d$ , $w \sim w_t$, and $w > w_t$. It can be noted that the PDFs have an approximately 
Gaussian peak around the initial height for all values of $w$, meaning that there is always a certain amount of samples that are simply not disturbed by the 
outlet. Apart from these samples, the great majority is completely discharged for $w > w_t$. This result is in consonance with that obtained earlier in a 
different context \cite{maga2}. In that study, it was shown that for large $w$, all blocked events occurred after only few grains had passed through the 
outlet. These facts indicate that the initial conditions may play an important role in jamming experiments. However, this issue needs further investigation.

\section{Conclusions}

Evidence of the jamming transition was observed in molecular dynamics simulations of open granular piles with rolling resistance, in consonance with the 
results found in granular piles without this interaction. This result strengthens the expectation that the transition exists in real granular piles and is 
probably affected by the system geometry. We observe that when there was rolling resistance, the piles built were more stable, denoted for the large mean height of the
samples, and also more robust against perturbations, since the critical aperture width increased. In future works, we plan to present a 
detailed study of the arching statistics for the two approaches considered here.

\begin{acknowledgements}
We are grateful for CNPq, FAPEMIG Brazilian funding agencies. APFA and GC thanks to CEFET-MG by the international interchange which made this interaction possible. 
\end{acknowledgements}


\begin{thebibliography}{50} 

\bibitem{antony}
S J Antony, W Hoyle, Y Ding, \textit{Granular materials: Fundamentals and applications}, The Royal Society of Chemistry, Cambridge (2004).

\bibitem{duran}
J Duran, \textit{Sands, powders and grains}, Springer, Berlin (1997).

\bibitem{halsey}
T Halsey, A Mehta, \textit{Challenges in granular physics}, World Scientific Publishing, New Jersey (2002).

\bibitem{yu} A Yu , K Dong , R Yang, S Luding, \textit{Powders and grains 2013: Proceedings of the 7th international conference on micromechanics of granular media}, AIP Series. Vol. 1542, Sydney, Australia (2013).

\bibitem{midi} GdR MiDi, Eur. Phys. J. E \textbf{14}, 341 (2004).

\bibitem{jaeger}
H M Jaeger, S R Nagel, \textit{Granular solids, liquids, and gases}, Rev. Mod. Phys. \textbf{68}, 1259 (1996).

\bibitem{cixous}
P Cixous, E Kolb, N Gaudouen, J-C Charme, \textit{Jamming and unjamming by penetration of a cylindrical intruder inside a 2 dimensional dense and disordered granular medium}, In: Powders and grains 2009, Proceedings of the 6th international conference on micromechanics of granular media \textbf{1145}, 539 (2009).

\bibitem{atman13}
A P F Atman, P Claudin, G Combe, R Mari, \textit{Mechanical response of an inclined frictional granular layer approaching unjamming}, Europhys. Lett. \textbf{101}, 44006 (2013).

\bibitem{liu}
A J Liu, S R Nagel, \textit{Jamming is not just cool any more}, Nature \textbf{396}, 21 (1998).

\bibitem{majmudar}
T S Majmudar, M Sperl, S Luding, R P Behringer, \textit{Jamming transition in granular systems}, Phys. Rev. Lett. \textbf{98}, 058001 (2007).

\bibitem{mankoc}
C Mankoc, A Janda, R Ar\'evalo, J M Pastor, I Zuriguel, A Garcimart\'i­n and D Maza. \textit{The flow rate of granular materials through an orifice}, Granul. Matter \textbf{9}, 407 (2007).

\bibitem{atman14}
A P F Atman, P Claudin, G Combe, G H B Martins, \textit{Mechanical properties of inclined frictional granular layers}, Granul. Matter \textbf{16}, 1 (2014).

\bibitem{katgert}
G Katgert, M van Hecke, \textit{Jamming and geometry of two-dimensional foams}, Europhys. Lett. \textbf{92}, 34002 (2010).

\bibitem{denkov}
N D Denkov, S Tcholakova, K Golemanov, A Lips, \textit{Jamming in sheared foams and emulsions, explained by critical instability of the films between neighboring bubbles and drops}, Phys. Rev. Lett. \textbf{103}, 118302 (2009).

\bibitem{kumar}
A Kumar, J Wu, \textit{Structural and dynamic properties of colloids near jamming transition}, Colloid. Surf. A \textbf{247}, 145151 (2004).

\bibitem{fluerasu}
A Fluerasu, A Moussaid, A Madsen, A Schofield, \textit{Slow dynamics and aging in colloidal gels studied by x-ray photon correlation spectroscopy}, Phys. Rev. E \textbf{76}, 010401 (2007).

\bibitem{duplantier}
B Duplantier, T C Halsey, V Rivasseau, \textit{Glasses and grains: Poincar\'e Seminar 2009}, Springer, Basel (2011).

\bibitem{hern1}
C S O'Hern, S A Langer, A J Liu, S R Nagel, \textit{Random packings of frictionless particles}, Phys. Rev. Lett. \textbf{88}, 075507 (2002).

\bibitem{hern2}
C O'Hern, L E Silbert, A J Liu, S R Nagel, \textit{Jamming at zero temperature and zero applied stress: The epitome of disorder}, Phys. Rev. E \textbf{68}, 011306 (2003).


\bibitem{manna}
S S Manna, D V Khakhar, \textit{Internal avalanches in a granular medium}, Phys. Rev. E \textbf{58}, R6935 (1998).

\bibitem{zuriguel}
I Zuriguel, A Garcimart\'in, D Maza, L A Pugnaloni, J M Pastor, \textit{Jamming during the discharge of granular matter from a silo}, Phys. Rev. E \textbf{71}, 051303 (2005).



\bibitem{to}
K To, P-Y Lai, H K Pak, \textit{Jamming of granular flow in a two-dimensional hopper}, Phys. Rev. Lett. \textbf{86}, 71 (2001).

\bibitem{garci}
A Garcimart\'in, I Zuriguel, L A Pugnaloni, A Janda, \textit{Shape of jamming arches in two-dimensional deposits of granular materials}, Phys. Rev. E \textbf{82}, 031306 (2010).

\bibitem{drescher}
A Drescher, A J Waters, C A Rhoades, \textit{Arching in hoppers .2. Arching theories and critical outlet size}, Powder Technol. \textbf{84}, 177 (1995).

\bibitem{maga2}
C F M Magalh\~aes, A P F Atman, J G Moreira, \textit{Segregation in arch formation}, Eur. J. Phys. E \textbf{35}, 38 (2012).

\bibitem{janda}
A Janda, I Zuriguel, A Garcimart\'in, L A Pugnaloni, D Maza, \textit{Jamming and critical outlet size in the discharge of a two-dimensional silo}, Europhys. Lett. \textbf{84}, 44002 (2008).

\bibitem{maga1}
C F M Magalh\~aes, J G Moreira, A P F Atman, \textit{Catastrophic regime in the discharge of a granular pile}, Phys. Rev. E \textbf{82}, 051303 (2010).

\bibitem{chevalier}
B Chevalier, G Combe, P Villard, \textit{Experimental and discrete element modeling studies of the trapdoor problem: Influence of the macro-mechanical frictional parameters}, Acta Geotech. \textbf{7}, 15 (2012). 

\bibitem{ai}
J Ai, J-F Chen, J M Rotter, J Y Ooi, \textit{Assessment of rolling resistance models in discrete element simulations}, Powder Technol. \textbf{206}, 269 (2011).

\bibitem{jiang}
M J Jiang, H-S Yu, D Harris, \textit{A novel discrete model for granular material incorporating rolling resistance}, Comput. Geotech. \textbf{32}, 340357 (2005).

\bibitem{zhou}
Y C Zhou, B D Wright, R Y Yang, B H Xu, A B Yu, \textit{Rolling friction in the dynamic simulation of sandpile formation}, Physica A \textbf{269}, 536 (1999).

\bibitem{iwashita}
K Iwashita, M Oda, \textit{Rolling resistance at contacts in simulation of shear band development by DEM}, J. Eng. Mech. \textbf{124}, 285292 (1998).

\bibitem{li}
X Li, X Chu, Y T Feng, \textit{A discrete particle model and numerical modeling of the failure modes of granular materials}, Eng. Computation. \textbf{22}, 894 (2005).

\bibitem{estrada}
N Estrada, E Az\'ema, F Radjai, A Taboada, \textit{Identification of rolling resistance as a shape parameter in sheared granular media}, Phys. Rev. E \textbf{84}, 011306 (2011).

\bibitem{charalampidou}
E-M Charalampidou, G Combe, G Viggiani, J Lanier, \textit{Mechanical behavior of mixtures of circular and rectangular 2D particles}, In: Powders and grains 2009: Proceedings of the 6th international conference on micromechanics of granular media, AIP Conf. Proc., Vol. 1145, Pag. 821, (2009).

\bibitem{rapaport}
D C Rapaport, \textit{The art of molecular dynamics simulation}, Cambridge University Press, Cambridge (2004).

\bibitem{swope}
W C Swope, H C Andersen, P H Berens, K R Wilson, \textit{Computer simulation method for the calculation of equilibrium constants for the formation of physical clusters of molecules: Application to small water clusters}, J. Chem. Phys. \textbf{76}, 637 (1982).

\bibitem{goldenberg}
C Goldenberg, A P F Atman, P Claudin, G Combe, I Goldhirsch, \textit{Scale separation in granular packings: Stress plateaus and fluctuations}, Phys. Rev. Lett. \textbf{96}, 168001 (2006).

\bibitem{pinto}
S F Pinto, A P F Atman, M S Couto, S G Alves, A T Bernardes, H F V Resende, E C Souza, \textit{Granular fingers on jammed systems: New fluidlike patterns arising in grain-grain invasion experiments}, Phys. Rev. Lett. \textbf{99}, 068001 (2007).

\bibitem{atman09}
A P F Atman, P Claudin, G Combe, \textit{Departure from elasticity in granular layers: Investigation of a crossover overload force}, Comput. Phys. Commun. \textbf{180}, 612 (2009).

\raggedbottom
\pagebreak

\bibitem{cundall}
P A Cundall, O D L Strack, \textit{A discrete numerical model for granular assemblies}, Geotechnique \textbf{29}, 47 (1979).

\bibitem{mindlin}
R D Mindlin, \textit{Compliance of elastic bodies in contact}, J. Appl. Mech. \textbf{71}, 259 (1949).

\bibitem{atman05}
A P F Atman, P Brunet, J Geng, G Reydellet, G Combe, P Claudin, R P Behringer, E Clement, \textit{Sensitivity of the stress response function to packing preparation}, J. Phys.: Cond. Matter \textbf{17}, S2391 (2005).


\bibitem{hertz}
H Hertz, \textit{On the contact of elastic solids}, J. Reine Angew. Math. \textbf{92}, 156 (1881).

\bibitem{allen}
M P Allen, D J Tildesley, \textit{Computer simulation of liquids}, Clarendon Press, Oxford (1987).



\bibitem{ref3} O O'Sullivan, \textit{Computing quaternions}, In: The art of numerical manipulation, Eds. A Q Rista, M Nadola, Pag. 132, North Holland, Amsterdam (2003).

\end{thebibliography}
\end{document}